\UseRawInputEncoding
\documentclass[10pt,aps,prl,twocolumn,superscriptaddress,postprint]{revtex4-2}

\usepackage[english]{babel}
\usepackage{physics}
\usepackage{graphicx}  
\usepackage{amssymb}
\usepackage{amsmath}  
\usepackage{dsfont}
\usepackage{multirow}
\usepackage{array}
\usepackage{booktabs}
\usepackage{stmaryrd}
\usepackage{mathrsfs}
\usepackage{tikz}
\usepackage{mathtools}
\usepackage{hyperref}
\usepackage[autostyle, english=american]{csquotes}
\usepackage{orcidlink}
\usepackage{xspace}
\usepackage{xcolor}
\MakeOuterQuote{"}

\newcommand{\iscat}[0]{i\textsc{scat}\xspace}
\newcommand{\ipsf}[0]{i\textsc{psf}\xspace}
\newcommand{\ipsfs}[0]{i\textsc{psf}s\xspace}
\newcommand{\rocs}[0]{\textsc{rocs}\xspace}

\addto\captionsenglish{%
}
\addto\captionsenglish{%
}

\newcommand{\figref}[2][{}]{Fig.~\hyperref[#2]{\ref{#2}#1}}
\newcommand{\tabref}[2][{}]{\tablename~\hyperref[#2]{\ref{#2}#1}}

\hypersetup{
  colorlinks = true,
  citecolor = blue,
  linkcolor = blue,
  urlcolor = black,
}

\begin{document}

\title{When Higher Resolution Reduces Precision: Quantum Limits of Off-Axis Interferometric Scattering Microscopy}
\author{Felix~\surname{Hitzelhammer}\,\orcidlink{0009-0006-5335-7363}}
\affiliation{Institute of Physics, NAWI Graz, \href{https://ror.org/01faaaf77}{\textcolor{blue}{University of Graz}}, 8010~Graz, Austria}
\email[Contact author:~]{felix.hitzelhammer@uni-graz.at}
\author{Jonathan~\surname{Dong}\,\orcidlink{0000-0002-8315-5571}}
\affiliation{Biomedical Imaging Group, \href{https://ror.org/02s376052}{\textcolor{blue}{ \'Ecole polytechnique fédérale de Lausanne}}, 1015 Lausanne, Switzerland\looseness=-1}
\author{Ulrich~\surname{Hohenester}\,\orcidlink{0000-0001-8929-2086}}
\affiliation{Institute of Physics, NAWI Graz, \href{https://ror.org/01faaaf77}{\textcolor{blue}{University of Graz}}, 8010~Graz, Austria}
\author{Thomas~\surname{Juffmann}\,\orcidlink{0000-0002-7098-5736}}
\affiliation{Faculty of Physics, VCQ, \href{https://ror.org/03prydq77}{\textcolor{blue}{University of Vienna}}, 1090 Vienna, Austria\looseness=-1}
\affiliation{Max Perutz Labs, {\href{https://ror.org/03prydq77}{\textcolor{blue}{University of Vienna}}},
1030 Vienna, Austria\looseness=-1}
\date{\today}
\begin{abstract}
 Coherent interferometric scattering microscopy (\iscat) enables nanoparticle tracking on microsecond timescales and with nanometer precision, and has become a key tool in structural and cellular biophysics. The achievable localization precision in such experiments is fundamentally limited by photon shot noise. Here, we analyze three-dimensional localization precision under oblique illumination in \iscat using the framework of (Quantum) Fisher Information. We show that tilting the illumination can enhance localization precision and accuracy per detected photon, while increasing robustness to defocusing. Surprisingly, rotating coherent scattering microscopy (\rocs), which incoherently averages oblique illuminations, achieves higher spatial resolution but lower localization precision. Our results establish the quantum limits of off-axis interferometric imaging and reveal that resolution and precision can behave in opposite ways --- a key insight for designing next-generation coherent microscopes.
\end{abstract}
\maketitle
\section{Introduction}
Elastic light scattering underpins diverse applications, from mass photometry~\cite{young2018} to fast particle tracking~\cite{taylor2019} and 3D imaging~\cite{kuppers2023}. A prominent technique is interferometric scattering microscopy (\iscat), which enhances contrast by interfering scattered light with a reference field from the partial reflection of wide-field incident illumination~\cite{taylor2019interferometric}. For 3D imaging, this geometry is particularly sensitive for axial localization~\cite{gholami2020point,dong2021}.
Recent advances suggest that oblique illumination can improve interferometric imaging. A key example is rotating coherent scattering microscopy (\rocs) \cite{ruh2018superior,junger2022100,iqbal2025enhanced}, where the angle of incidence rotates around the optical axis. \rocs offers higher spatial resolution than on-axis schemes. 

In coherent imaging, measurement precision is limited by shot noise, i.e., quantum fluctuations in detected photons. While collecting more photons increases precision, this is constrained by sample dynamics, camera well depth, photodamage, or setup stability. It is therefore crucial to maximize the information extracted per photon. Two quantities capture this shot-noise-limited precision. The Fisher information (FI) quantifies how much information Poisson-distributed measurements carry about a parameter~\cite{van2004detection,backlund2018fundamental}, and is bounded above by the quantum Fisher information (QFI), which describes the information in the quantum state of the scattered light~\cite{bouchet2021maximum}. Their inverses define the Cramér--Rao bound (CRB) and quantum Cramér--Rao bound (QCRB), which set limits on localization precision based on the noisy measurements, or based on the quantum state itself, respectively.

Previous work~\cite{dong2021} computed FI and QFI in conventional \iscat at normal incidence, revealing a drastic difference between axial and transverse localization. However, the off-axis case remains unexplored, as it requires modeling of non-paraxial field propagation and of scatterer-interface couplings.

In this paper, we investigate how off-axis illumination affects localization precision in coherent scattering. Using a numerical toolbox for coherent imaging~\cite{hitzelhammer24}, we simulate electric fields and compute classical and quantum FI with associated CRBs. We first demonstrate how the spatial distribution of the FI flow~\cite{hupfl_continuity_2024} depends on illumination angle and how this influences how much information can be gathered with a certain numerical aperture ($\mathrm{NA}$). We then show that oblique illumination increases the QFI of the scattered light and yields up to twofold improvements in transverse localization in \iscat. Lastly, we analyze CRBs in \rocs, which has been demonstrated to improve spatial resolution in interferometric imaging. While \rocs uses oblique illumination to achieve this, it incoherently integrates images obtained under various illumination angles. Our information-theoretical analysis reveals that this reduces localization precision, despite the improved spatial resolution.
\begin{figure*}
    \centering
    \hspace*{-0.6cm}
    \includegraphics[scale=0.6]{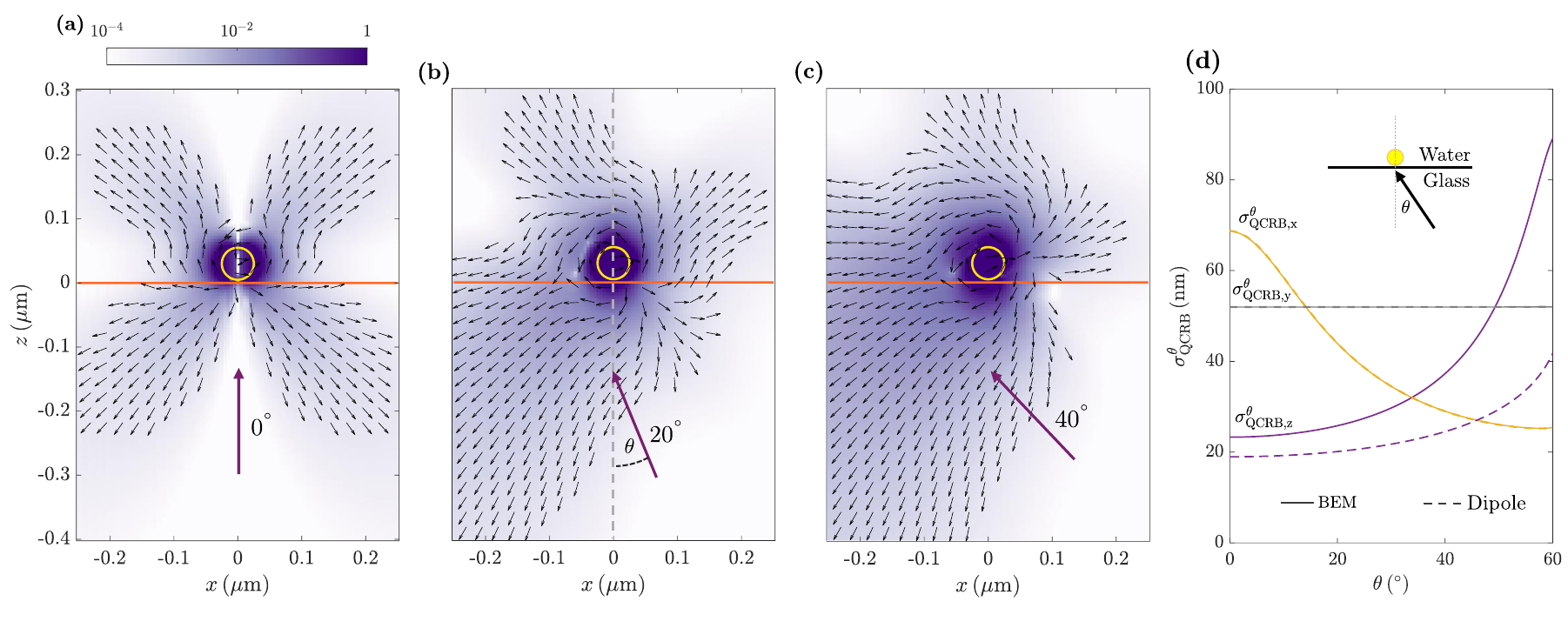}
    \caption{Fisher information (FI) flow (a.u.) in the $x$-$z$ plane for a gold nanosphere (golden circle) near a glass--water interface (orange line), together with the quantum Cramér--Rao bounds (QCRBs) for 3D localization precision under off-axis illumination. (a) FI flow under on-axis illumination, i.e.\ along the optical axis (gray dashed line). (b) FI flow for an illumination angle of $20^{\circ}$. (c) FI flow for an illumination angle of $40^{\circ}$. (d) QCRBs for localization precision along the $x$ (orange line), $y$ (grey line), and $z$ (purple line) directions. Table~S1 in the supporting information reports the measurement-independent gain in localization precision predicted by the QCRBs for off-axis illumination in panel (d).}
    \label{fig1:FI flow}
\end{figure*}
\section{Results}
To quantify the effects of illumination geometry on localization precision, we model the detected intensity as the coherent superposition of scattered and reflected fields~\cite{ginsberg2025interferometric}
\begin{align}\label{Intensity distribution}
   I^{(\theta,z_{\mathrm{f}})}(x,y) &= | \mathbf{E}_{\mathrm{s}}e^{i\phi_{\mathrm{s}}} + \mathbf{E}_{\mathrm{r}}e^{i\phi_{\mathrm{r}}} |^{2} \notag \\ 
   &= | \mathbf{E}_{\mathrm{s}}|^{2} + | \mathbf{E}_{\mathrm{r}}|^{2} + 2\,E_{\mathrm{s}} E_{\mathrm{r}} \cos  \phi\,,
\end{align}
\noindent with lateral coordinates $(x,y)$, incident angle $\theta$, defocus $z_{\mathrm{f}}$, and relative phase $\phi=\phi_{\mathrm{s}}-\phi_{\mathrm{r}}$. For brevity, the $(x,y)$-dependence is omitted on the right-hand side. The interferometric point-spread function (\ipsf) under off-axis illumination is computed using a numerical toolbox~\cite{hitzelhammer24} that combines the boundary element method for scattering fields~\cite{hohenester2022nanophotonic,hohenester2024nanophotonic} with a vectorial high-NA imaging model~\cite{dong2021,liu2025revisiting}. We compare our numerical results to an analytic simplified dipole model, denoted as dipole approximation, that uses the quasistatic approximation and models the nanosphere as a small polarizable sphere~\cite{gholami2020point}. We then consider the far-fields of the resulting dipole emitter in the vicinity of the substrate~\cite{novotny2012principles}.
We use the results of these simulations to calculate the QFI, the FI flow, and the QCRBs for localization precision, as detailed in the Methods section. These quantities are calculated per scattered photon, since the scattered photons carry the information about the particle location.
\\
\begin{figure*}
    \centering
    \hspace*{-1.5cm}
    \includegraphics[scale=0.7]{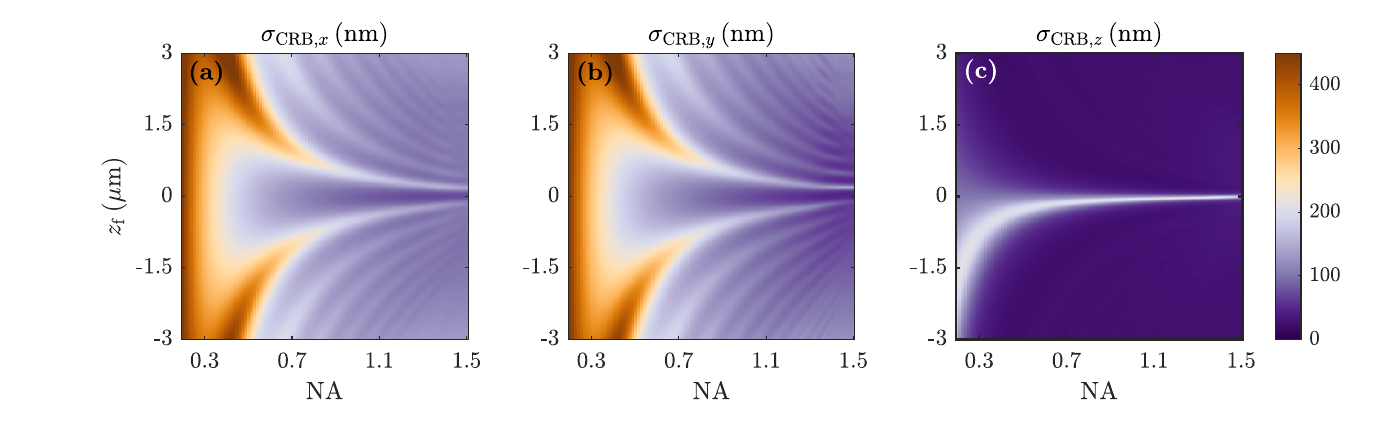}    \caption{Cramér--Rao bounds (CRBs) on the standard deviation of 3D localization precision for a gold nanosphere near a glass-water interface at $\theta = 0^{\circ}$ over a large defocus range and varying numerical aperture ($\mathrm{NA}$). Panels (a--c) show the CRB with respect to the $x$-, $y$-, and $z$-coordinates, respectively. The colorbar represents the value of the CRBs in $\mathrm{nm}$.}
    \label{fig2: Optimum NA}
\end{figure*}
\begin{figure*}
    \centering
    \hspace*{-0.5cm}
   \includegraphics[scale=0.8]{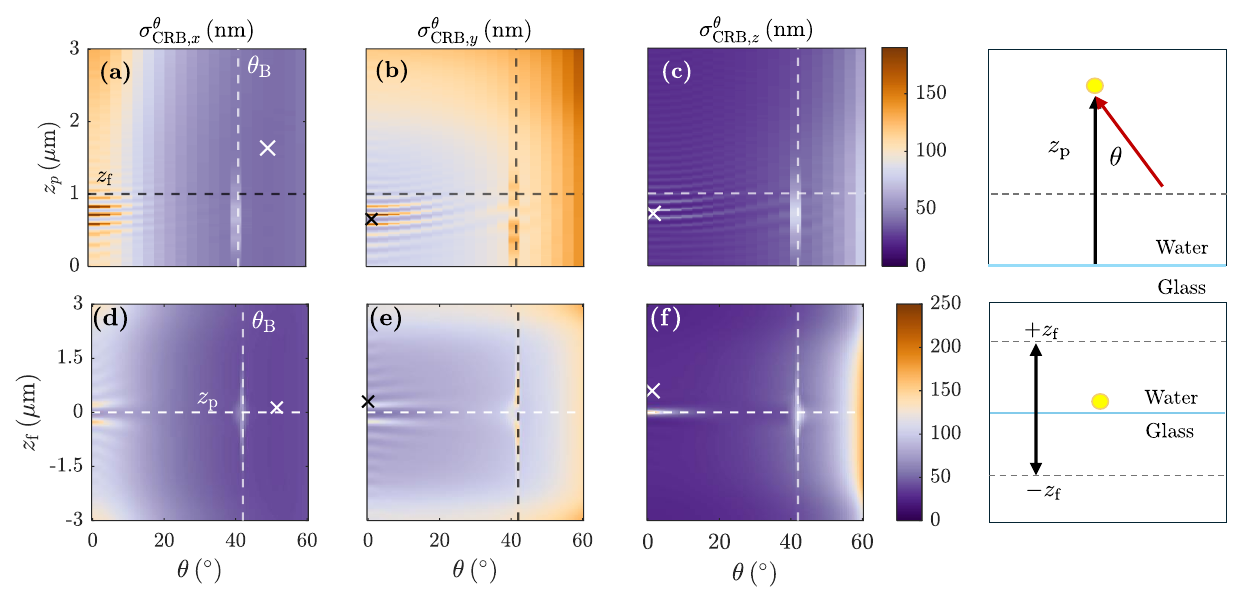}
    \caption{Cramér--Rao bounds (CRBs) on the standard deviation of 3D localization precision for a gold nanosphere near a glass-water interface under off-axis illumination. In panels (a--c), results are shown for a fixed focus plane $z_{\mathrm{f}}=1\,\mathrm{\mu m}$ (dashed line), while the particle position $z_{\mathrm{p}}$ is varied from near the interface up to $3\, \mu\mathrm{m}$. In panels (d--f), the particle position $z_{\mathrm{p}}$ is fixed near the interface (dashed line), and $z_{\mathrm{f}}$ is varied over a large defocus range. The vertical dashed lines in panels (a--f) indicate the Brewster angle $\theta_{\mathrm{B}} \approx 42^{\circ}$. Furthermore, in panels (a--f) the global minimum is denoted by a cross, indicating the best possible bound. Each colorbar represents the CRBs, with values given in $\mathrm{nm}$.}
     \label{fig3: CRBS_zp_zf}
\end{figure*}
\begin{figure*}
    \centering
    \hspace*{-0.4cm}
    \includegraphics[scale=0.72]{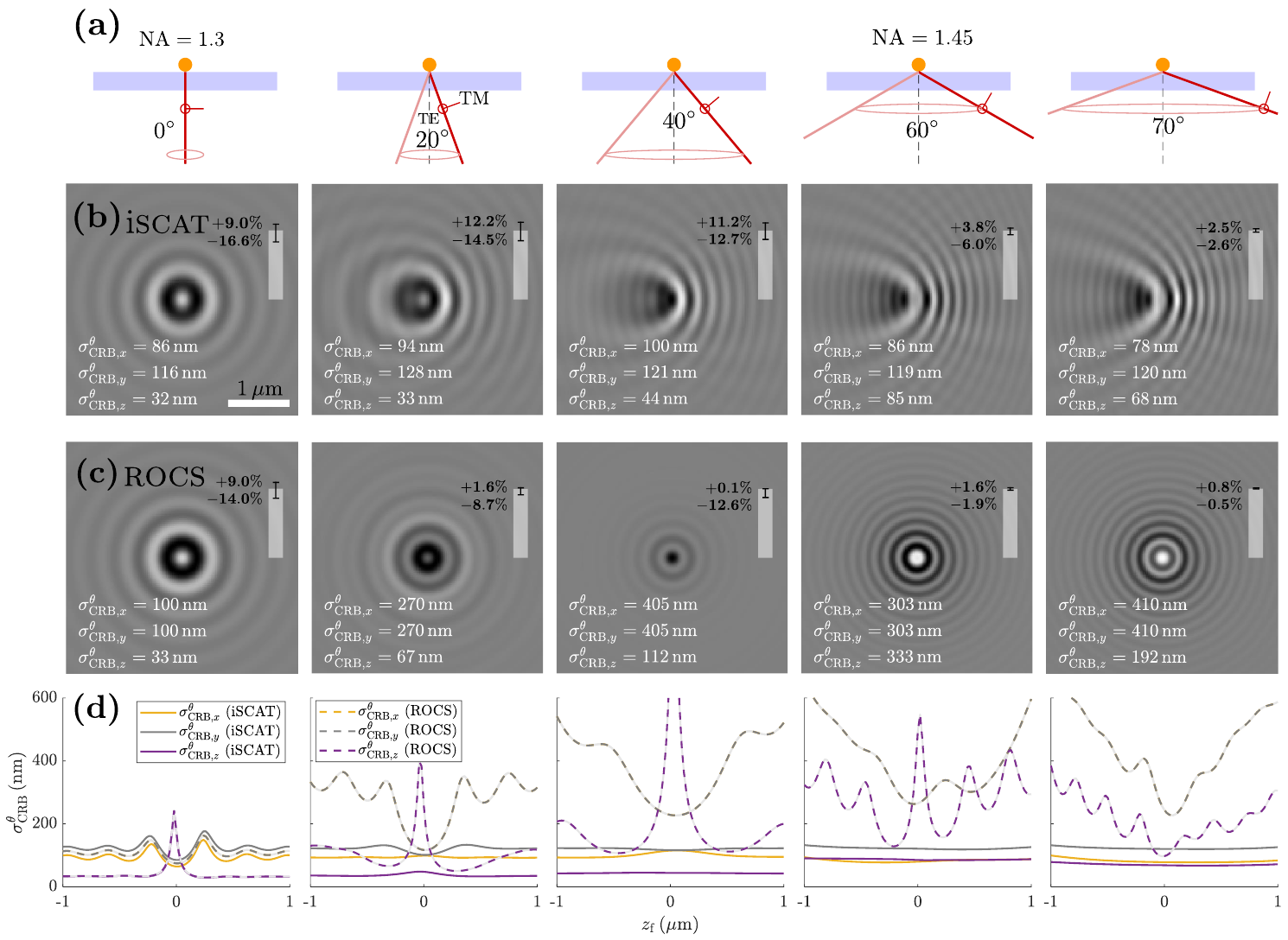}
    \caption{Interferometric scattering microscopy (\iscat) and rotating coherent scattering microscopy (\rocs) images for a gold nanosphere on top of a glass substrate. (a) We consider an incoming plane wave with TE polarization and the different angles reported in the insets. (b) \iscat images for a focus position of $z_{\mathrm{f}}=500\,\mathrm{nm}$. (c) Corresponding \rocs images, which we obtain by rotating the incoming wave around the $z$-axis and summing (incoherently) over the resulting \iscat images. The Cramér--Rao bounds (CRBs) obtained from the \iscat and \rocs images are reported in the respective panels, where $\sigma_{x}$ corresponds to $\sigma^{\theta}_{{\mathrm{CRB}},x}$ with corresponding assignments for $y$ and $z$. 
    (d) The CRBs as a function of defocus show that the localization precision in \rocs is consistently worse than in oblique \iscat. Since the CRBs for $x$ and $y$ in \rocs are identical, they are represented by a single gray dashed line.}
    \label{fig4:roc_vs_iscat}
\end{figure*}
\\
\noindent
\textbf{Fisher information flow}\\
Panels (a--c) of Fig.~\ref{fig1:FI flow} show simulation results of the FI flow for a gold nanosphere near a glass--water interface under off-axis illumination. The particle, illuminated from below by a TM-polarized plane wave at $520\,\mathrm{nm}$ and angle $\theta$ relative to the optical axis $z$, acts as the source of the flow. For on-axis illumination ($\theta=0$), Fig.~\ref{fig1:FI flow}(a) shows that the forward- and backward-scattered fields carry nearly equal information. The flow propagates at an angle with respect to the optical axis, highlighting the role of high-NA detection. With off-axis illumination, the flow redistributes and is enhanced in the backward direction, as revealed in panels (b) and (c) of Fig.~\ref{fig1:FI flow}, suggesting that more information can be collected in \iscat, where the imaging objective gathers backscattered light.

\noindent
\textbf{Dependence on numerical aperture}\\ 
We quantify this conjecture in Fig.~\ref{fig1:FI flow}(d), where the QCRBs for 3D localization precision as a function of the illumination angle $\theta$ are shown.
The solid lines are obtained using the \textsc{nanobem} simulation toolbox, while the dashed lines are calculated within the dipole approximation. We observe that the QCRBs are strongly dependent on the illumination geometry. Specifically, we find that the QCRBs in the $x$-direction improve by a factor of $\approx2.8$, that is, they decrease with illumination angle, while those in the $y$-direction remain almost constant. This is a consequence of the incoming beam being tilted in the $x$-$z$ plane, leading to more information being collected about the $x$-coordinate, in agreement with panels (a--c) of Fig.~\ref{fig1:FI flow}. Interestingly, for $z$-localization precision, the QCRBs worsen at the glass--water interface, as shown in Fig.~\ref{fig1:FI flow}(d), but improve at the glass--air interface, as demonstrated in Figure~S1(d) of the Supporting Information (SI).
The measurement-independent improvements achievable with oblique illumination are summarized in Table~S1 (SI). In both cases, Fig.~\ref{fig1:FI flow}(d) and Figure~S1(d) show that the full BEM calculation yields different results from the dipole approximation. This discrepancy is particularly evident in Figure~S1(d), where the dipole model predicts a peak at the critical angle that is absent in the BEM calculation. A closer analysis reveals that this difference arises from higher-order multipoles induced in the sphere near the glass substrate, which are neglected in the dipole approximation. We note that the discrepancy would be smaller for non-metallic nanospheres.

For on-axis illumination, the FI flow in Fig.~\ref{fig1:FI flow}(a) indicates that 3D localization precision depends strongly on the $\mathrm{NA}$ of the imaging system.
This is confirmed in Fig.~\ref{fig2: Optimum NA}, which shows CRBs versus $\mathrm{NA}$ over a large defocus range. As expected, in panels (a--c) the CRBs decrease with increasing $\mathrm{NA}$ , but the deterioration is only a factor of $\approx2.3$ when comparing $\mathrm{NA}=0.5$ to $\mathrm{NA}=1.5$ near $z_{\mathrm{f}}=0\,\mathrm{nm}$. This suggests the feasibility of low-cost \iscat setups with modest $\mathrm{NA}$, provided samples are sparse to minimize interference from nearby scatterers. We also note the excellent axial CRBs in Fig.~\ref{fig2: Optimum NA}(c), arising from the strong axial dependence of the \iscat signal on the particle–substrate distance $z_p$~\cite{dong2021}. \\
\\
\noindent
\textbf{Influence of defocus and particle position}\\ 
We next analyze the behavior of the CRBs under variations of $(\theta,z_{\mathrm{f}})$ and $(\theta,z_{\mathrm{p}})$, where $z_{\mathrm{f}}$ and $z_{\mathrm{p}}$ denote the focus plane and particle position, respectively. Figure~\ref{fig3: CRBS_zp_zf} shows 3D localization CRBs at a glass--water interface for both parameter sets in an \iscat geometry. In panels (a--c), the defocus plane is fixed at $1~\mathrm{\mu m}$ (vertical dashed line) while $z_{\mathrm{p}}$ is varied, a configuration relevant to nanoparticle tracking on cell membranes. In panels (d--f), the particle is near the interface and $z_{\mathrm{f}}$ is varied. The localization precision in $x$ improves with illumination angle, as shown in panels (a) and (d), consistent across the entire defocus and particle range. The improved QCRBs from Fig.~\ref{fig1:FI flow}(d) translate into improved CRBs in the \iscat geometry, both for the glass--water interface (see panels (a) and (d) of Fig.~\ref{fig3: CRBS_zp_zf}) and for the glass--air interface (see Figure~S1(a)).
In contrast, $\sigma^{\theta}_{\mathrm{CRB},y}$ worsens with increasing $\theta$, as shown in panels (b) and (e) of Fig.~\ref{fig3: CRBS_zp_zf}, and $\sigma^{\theta}_{\mathrm{CRB},z}$ generally worsens at larger $\theta$, as illustrated in panels (c) and (f). For a glass–air interface, however, axial precision improves with $\theta$, as shown in the panels (c) and (d) of Figure~S1. Interestingly, although Brewster-angle illumination increases contrast~\cite{hitzelhammer24}, the CRBs in all three spatial directions exhibit local maxima, corresponding to reduced localization precision. This underscores that contrast alone is not a reliable indicator. Furthermore, by comparing optimal bounds for off-axis illumination, denoted by a cross in Fig.~\ref{fig3: CRBS_zp_zf}(d), with an on-axis geometry, we report an enhancement factor of $\approx1.8$ for $x$ at the glass--water interface. A similar analysis versus particle position in Fig.~\ref{fig3: CRBS_zp_zf}(a) shows a gain of $\approx1.9$ in $x$. For clarity, panels (a) and (b) in Figure~S2 show these enhancements for the glass--water interface. For the glass--air interface in Figure S1(c), the global minima (marked by crosses) correspond to an enhancement factor of $\approx1.8$ in $z$. Panels (c) and (d) in Figure~S2 further show this gain across the entire defocus range.

At low illumination angles, the CRBs oscillate with $z_f$, as can be seen in panels (a--c) of Fig~\ref{fig3: CRBS_zp_zf} and Figure~S2, due to interference between scattered and reference light~\cite{dong2021}. These oscillations disappear at higher angles. Similarly, on-axis CRBs oscillate with $z_p$, but vanish under oblique illumination. Off-axis illumination can thus enable more robust particle localization, while maintaining precision. This can be understood qualitatively by recognizing that, for on-axis and $z_p \approx z_f$, most FI is concentrated at the \ipsf center, whose contrast oscillates rapidly with the phase between scattered and reflected light, leading to oscillations in the FI and CRBs~\cite{dong2021}. When the particle is out of focus, FI shifts to the outer rings of the \ipsf, which move slightly with $z_p$ but leave the integrated FI nearly constant. Oblique illumination shows a similar behavior: oscillatory structure near the \ipsf center but smoother integrated FI, as depicted in Fig.~\ref{fig4:roc_vs_iscat}(b). \\
\\
\noindent
\textbf{Comparison with rotating coherent scattering microscopy}\\ Lastly, we examine whether the improved CRBs under oblique illumination extend to rotating coherent scattering microscopy (\rocs), where oblique illumination is rotated around the optical axis and images are summed incoherently. Figure~\ref{fig4:roc_vs_iscat} compares the \ipsfs for \iscat and \rocs, shown in panels (b) and (c), respectively, at $z_f = 500\,\mathrm{nm}$. While \rocs yields a narrower \ipsf and thus higher spatial resolution~\cite{ruh2018superior,iqbal2025enhanced}, its CRBs for 3D localization ($\sigma^{\theta}_{\mathrm{CRB},\gamma_{i}}$) are significantly worse, see values in panels (b),(c). Incoherent averaging suppresses the outer interference rings, leading to loss of information across all defocus values (panel d). For example, at $\theta=60^{\circ}$, $\sigma_{\mathrm{CRB},x}$ in \rocs is more than twice that in \iscat, requiring $\sim$5× more photons for similar precision. In practice, dense samples complicate this trade-off: uncontrolled interference degrades \iscat precision, whereas \rocs remains more robust under such conditions.

\section{Discussion}
We have discussed 3D localization precision in interferometric microscopy with oblique illumination. Based on a precise computation of (Quantum) Fisher Information, we find that oblique illumination can increase both the transverse and axial localization precision by roughly a factor of two. Alternatively, at a given precision, measurements could be done four times faster, or at four times lower excitation intensity. The former can be crucial when imaging fast protein dynamics ~\cite{foley_mass_2021, heermann_mass-sensitive_2021, bujak2021}, the latter when imaging live cells, which often show functional changes when illuminated with light, even without exogenous fluorophores and at moderate photon fluence~\cite{lipovsky_visible_2010, abana_characterization_2017, Cuny2022LiveCM}. The framework of FI flow provides an intuitive explanation for this phenomenon and also visualizes the dependence of the achievable localization precision on the system's numerical aperture. Beyond improving localization precision, off-axis illumination also enhances robustness in axial localization. 

When comparing \iscat and \rocs imaging modalities, we find that \rocs provides worse localization precision, despite its higher spatial resolution. Interestingly, both the higher spatial resolution and the worsening of the localization precision have their origin in the incoherent integration of images from different illumination directions. 
A similarly counterintuitive result is found for \iscat with Brewster angle illumination. Here, the contrast improves~\cite{hitzelhammer24}, but the CRBs for localization precision worsen. These examples show that resolution and contrast alone are not sufficient to fully describe the performance of an imaging modality. 

These findings illustrate how quantum information principles can guide the design of optical imaging modalities, with implications for high-speed, low-intensity microscopy of biological specimens. As an example, with respect to  \rocs, it would be advantageous to record one image per illumination angle. While the summed-up image provides high spatial resolution and convenient suppression of unwanted interference - key advantages of \rocs - the individual frames hold enhanced Fisher Information that can enable high-precision tracking. While data recording would be slower than in traditional \rocs, the approach could still yield sub-ms temporal resolution as camera technology reaches frame rates of tens of kHz~\cite{taylor2019}.

Beyond applications in bioimaging, our results can be generalized to any imaging scheme based on coherent scattering, potentially benefiting plasmonics~\cite{schmid2014plasmonic}, nanoparticle cooling and trapping~\cite{levitation21}, and the imaging of ultracold atoms~\cite{ketterle1999making}. 

\section{Methods}
\noindent
\textbf{(Q)FI and (Q)CRB calculations} \\
The CRB sets a lower limit on the variance of any unbiased estimator $\hat{\gamma}$~\cite{van2004detection}. In the many-photon limit the maximum likelihood estimator is unbiased and efficient, making the CRB directly relevant for localization precision in interferometric imaging. The CRB is given by the inverse of the FI, which quantifies the sensitivity of the data to changes in parameters~\cite{shechtman2014}. We estimate the 3D localization precision for $\gamma=(x,y,z)$ using the FI matrix
\begin{equation}\label{eq:fisher information matrix}
    [\boldsymbol{\mathcal{I}}(\gamma)]_{ij} = \int_{C} dx\,dy \, \frac{\left[\partial_{\gamma_{i}} I^{(\theta,z_{\mathrm{f}})}(x,y)\right] \left[\partial_{\gamma_{j}} I^{(\theta,z_{\mathrm{f}})}(x,y)\right]}{ I^{(\theta,z_{\mathrm{f}})}(x,y) } \, ,
\end{equation}
where $C$ denotes the region of integration across the image plane. We note that the form of the FI matrix depends on the detection statistics, which here are Poissonian, corresponding to the shot-noise-limited regime~\cite{Chao2016}. Since only few parameters are estimated, the derivatives in Eq.~\eqref{eq:fisher information matrix} are computed via finite differences. The FI matrix bounds the variance component-wise~\cite{Chao2016,dong2021}
\begin{equation}\label{eq:CR_inequality_component}
\mathrm{Var}(\hat{\gamma}_{j}) \geq [\boldsymbol{\mathcal{I}}^{-1}(\gamma)]_{jj} \geq \frac{1}{[\boldsymbol{\mathcal{I}}(\gamma)]_{jj}}\,,
\end{equation}
where the second inequality gives a weaker bound. This applies to localization precision when assessing the sensitivity of the \ipsf to a single parameter $\gamma_j$, assuming all others are perfectly known. The corresponding CRB on the standard deviation is then given by
\begin{equation}\label{eq:CRB}
    \sigma_{\mathrm{CRB}, \gamma_j} = \frac{1}{\sqrt{[\boldsymbol{\mathcal{I}}(\gamma)]_{jj}}} \,.
\end{equation}
In contrast to the classical CRB, the
QCRB is measurement-independent and thus provides the ultimate benchmark in evaluating the potential of an imaging geometry~\cite{backlund2018fundamental, juffmann2025quantum}.
In~\cite{bouchet2021maximum} an analytical expression of the QFI in the context of coherent scattering is derived, which is applicable to \iscat where the scattered and reflected fields interfere coherently. Building on this, we use the representation of the scattered light as a superposition of coherent states with amplitudes $\epsilon(\Theta, \phi)$, leading to the following expression for the QFI~\cite{dong2021}
\begin{equation}\label{eq:QFI}
    \mathcal{K}_{jj} = 4 \int_{0}^{\alpha} \,d\Theta \int_{0}^{2 \pi} d\phi | \partial_{j} \epsilon(\Theta, \phi)|^{2} \, ,
\end{equation}
where $\alpha$ is the maximal
angular aperture of the objective. 
We again use finite differences to compute partial derivatives and normalize the state to one scattered photon going through the backfocal plane~\cite{dong2021}, i.e. $\int_{0}^{\alpha} \,d\Theta \int_{0}^{2 \pi} d\phi | \epsilon(\Theta, \phi)|^{2} = 1$.
We can formulate a quantum counterpart to Eq.~\eqref{eq:CRB} as
\begin{equation}\label{eq:QCRB}
    \sigma_{\mathrm{QCRB},{\gamma_{j}}} = \frac{1}{\sqrt{\mathcal{K}_{jj}}}  \, .
\end{equation}
Recent work has introduced the flow of FI in electromagentic scattering problems~\cite{hupfl_continuity_2024}. In the context of \iscat, the FI flow regarding localization precision in the $x$-direction is given by
\begin{equation}\label{eq:FI-flow}
\mathbf{S}^{\mathrm{FI}}_{x}=\frac 2{\hbar\omega}\Re{\partial_{x}\mathbf{E}_{\rm sca}\times\partial_{x}\mathbf{H}^{*}_{sca}}\,,
\end{equation}
where $\mathbf{E}_{\rm sca}$, $\mathbf{H}_{\rm sca}$ denote the fields scattered by the particle and the derivatives $\partial_x$ are taken with respect to the particle position. 
\section{Acknowledgments}
This project has received funding from the European Research Council (ERC) under the European Union’s Horizon 2020 research and innovation program (Grant Agreement No. 758752) and the Swiss National Science Foundation (Grant PZ00P2\_216211). The authors acknowledge the financial support by the University of Graz.
\subsection{Data availability}
The data and simulation scripts supporting the findings of this study are available from the corresponding author upon reasonable request.
\subsection{Competing interests}
The authors declare no competing interests.
\bibliographystyle{apsrev4-2}
\bibliography{references}

\begin{thebibliography}{31}%
\makeatletter
\providecommand \@ifxundefined [1]{%
 \@ifx{#1\undefined}
}%
\providecommand \@ifnum [1]{%
 \ifnum #1\expandafter \@firstoftwo
 \else \expandafter \@secondoftwo
 \fi
}%
\providecommand \@ifx [1]{%
 \ifx #1\expandafter \@firstoftwo
 \else \expandafter \@secondoftwo
 \fi
}%
\providecommand \natexlab [1]{#1}%
\providecommand \enquote  [1]{``#1''}%
\providecommand \bibnamefont  [1]{#1}%
\providecommand \bibfnamefont [1]{#1}%
\providecommand \citenamefont [1]{#1}%
\providecommand \href@noop [0]{\@secondoftwo}%
\providecommand \href [0]{\begingroup \@sanitize@url \@href}%
\providecommand \@href[1]{\@@startlink{#1}\@@href}%
\providecommand \@@href[1]{\endgroup#1\@@endlink}%
\providecommand \@sanitize@url [0]{\catcode `\\12\catcode `\$12\catcode `\&12\catcode `\#12\catcode `\^12\catcode `\_12\catcode `\%12\relax}%
\providecommand \@@startlink[1]{}%
\providecommand \@@endlink[0]{}%
\providecommand \url  [0]{\begingroup\@sanitize@url \@url }%
\providecommand \@url [1]{\endgroup\@href {#1}{\urlprefix }}%
\providecommand \urlprefix  [0]{URL }%
\providecommand \Eprint [0]{\href }%
\providecommand \doibase [0]{https://doi.org/}%
\providecommand \selectlanguage [0]{\@gobble}%
\providecommand \bibinfo  [0]{\@secondoftwo}%
\providecommand \bibfield  [0]{\@secondoftwo}%
\providecommand \translation [1]{[#1]}%
\providecommand \BibitemOpen [0]{}%
\providecommand \bibitemStop [0]{}%
\providecommand \bibitemNoStop [0]{.\EOS\space}%
\providecommand \EOS [0]{\spacefactor3000\relax}%
\providecommand \BibitemShut  [1]{\csname bibitem#1\endcsname}%
\let\auto@bib@innerbib\@empty
\bibitem [{\citenamefont {Young}\ \emph {et~al.}(2018)\citenamefont {Young}, \citenamefont {Hundt}, \citenamefont {Cole}, \citenamefont {Fineberg}, \citenamefont {Andrecka}, \citenamefont {Tyler}, \citenamefont {Olerinyova}, \citenamefont {Ansari}, \citenamefont {Marklund}, \citenamefont {Collier}, \citenamefont {Chandler}, \citenamefont {Tkachenko}, \citenamefont {Allen}, \citenamefont {Crispin}, \citenamefont {Billington}, \citenamefont {Takagi}, \citenamefont {Sellers}, \citenamefont {Eichmann}, \citenamefont {Selenko}, \citenamefont {Frey}, \citenamefont {Riek}, \citenamefont {Galpin}, \citenamefont {Struwe}, \citenamefont {Benesch},\ and\ \citenamefont {Kukura}}]{young2018}%
  \BibitemOpen
  \bibfield  {author} {\bibinfo {author} {\bibfnamefont {G.}~\bibnamefont {Young}}, \bibinfo {author} {\bibfnamefont {N.}~\bibnamefont {Hundt}}, \bibinfo {author} {\bibfnamefont {D.}~\bibnamefont {Cole}}, \bibinfo {author} {\bibfnamefont {A.}~\bibnamefont {Fineberg}}, \bibinfo {author} {\bibfnamefont {J.}~\bibnamefont {Andrecka}}, \bibinfo {author} {\bibfnamefont {A.}~\bibnamefont {Tyler}}, \bibinfo {author} {\bibfnamefont {A.}~\bibnamefont {Olerinyova}}, \bibinfo {author} {\bibfnamefont {A.}~\bibnamefont {Ansari}}, \bibinfo {author} {\bibfnamefont {E.~G.}\ \bibnamefont {Marklund}}, \bibinfo {author} {\bibfnamefont {M.~P.}\ \bibnamefont {Collier}}, \bibinfo {author} {\bibfnamefont {S.~A.}\ \bibnamefont {Chandler}}, \bibinfo {author} {\bibfnamefont {O.}~\bibnamefont {Tkachenko}}, \bibinfo {author} {\bibfnamefont {J.}~\bibnamefont {Allen}}, \bibinfo {author} {\bibfnamefont {M.}~\bibnamefont {Crispin}}, \bibinfo {author} {\bibfnamefont {N.}~\bibnamefont {Billington}}, \bibinfo {author} {\bibfnamefont
  {Y.}~\bibnamefont {Takagi}}, \bibinfo {author} {\bibfnamefont {J.~R.}\ \bibnamefont {Sellers}}, \bibinfo {author} {\bibfnamefont {C.}~\bibnamefont {Eichmann}}, \bibinfo {author} {\bibfnamefont {P.}~\bibnamefont {Selenko}}, \bibinfo {author} {\bibfnamefont {L.}~\bibnamefont {Frey}}, \bibinfo {author} {\bibfnamefont {R.}~\bibnamefont {Riek}}, \bibinfo {author} {\bibfnamefont {M.~R.}\ \bibnamefont {Galpin}}, \bibinfo {author} {\bibfnamefont {W.~B.}\ \bibnamefont {Struwe}}, \bibinfo {author} {\bibfnamefont {J.~L.~P.}\ \bibnamefont {Benesch}},\ and\ \bibinfo {author} {\bibfnamefont {P.}~\bibnamefont {Kukura}},\ }\href {https://doi.org/10.1126/science.aar5839} {\bibfield  {journal} {\bibinfo  {journal} {Science}\ }\textbf {\bibinfo {volume} {360}},\ \bibinfo {pages} {423} (\bibinfo {year} {2018})}\BibitemShut {NoStop}%
\bibitem [{\citenamefont {Taylor}\ \emph {et~al.}(2019)\citenamefont {Taylor}, \citenamefont {Mahmoodabadi}, \citenamefont {Rauschenberger}, \citenamefont {Giessl}, \citenamefont {Schambony},\ and\ \citenamefont {Sandoghdar}}]{taylor2019}%
  \BibitemOpen
  \bibfield  {author} {\bibinfo {author} {\bibfnamefont {R.~W.}\ \bibnamefont {Taylor}}, \bibinfo {author} {\bibfnamefont {R.~G.}\ \bibnamefont {Mahmoodabadi}}, \bibinfo {author} {\bibfnamefont {V.}~\bibnamefont {Rauschenberger}}, \bibinfo {author} {\bibfnamefont {A.}~\bibnamefont {Giessl}}, \bibinfo {author} {\bibfnamefont {A.}~\bibnamefont {Schambony}},\ and\ \bibinfo {author} {\bibfnamefont {V.}~\bibnamefont {Sandoghdar}},\ }\href {https://doi.org/10.1038/s41566-019-0414-6} {\bibfield  {journal} {\bibinfo  {journal} {Nat. Photonics}\ }\textbf {\bibinfo {volume} {13}},\ \bibinfo {pages} {480} (\bibinfo {year} {2019})}\BibitemShut {NoStop}%
\bibitem [{\citenamefont {Küppers}\ \emph {et~al.}(2023)\citenamefont {Küppers}, \citenamefont {Albrecht}, \citenamefont {Kashkanova}, \citenamefont {Lühr},\ and\ \citenamefont {Sandoghdar}}]{kuppers2023}%
  \BibitemOpen
  \bibfield  {author} {\bibinfo {author} {\bibfnamefont {M.}~\bibnamefont {Küppers}}, \bibinfo {author} {\bibfnamefont {D.}~\bibnamefont {Albrecht}}, \bibinfo {author} {\bibfnamefont {A.~D.}\ \bibnamefont {Kashkanova}}, \bibinfo {author} {\bibfnamefont {J.}~\bibnamefont {Lühr}},\ and\ \bibinfo {author} {\bibfnamefont {V.}~\bibnamefont {Sandoghdar}},\ }\href {https://doi.org/10.1038/s41467-023-37497-7} {\bibfield  {journal} {\bibinfo  {journal} {Nat. Commun.}\ }\textbf {\bibinfo {volume} {14}},\ \bibinfo {pages} {1962} (\bibinfo {year} {2023})}\BibitemShut {NoStop}%
\bibitem [{\citenamefont {Taylor}\ and\ \citenamefont {Sandoghdar}(2019)}]{taylor2019interferometric}%
  \BibitemOpen
  \bibfield  {author} {\bibinfo {author} {\bibfnamefont {R.~W.}\ \bibnamefont {Taylor}}\ and\ \bibinfo {author} {\bibfnamefont {V.}~\bibnamefont {Sandoghdar}},\ }\bibinfo {title} {{Interferometric Scattering (iSCAT) Microscopy and Related Techniques}},\ in\ \href {https://doi.org/10.1007/978-3-030-21722-8_2} {\emph {\bibinfo {booktitle} {Label-Free Super-Resolution Microscopy}}}\ (\bibinfo  {publisher} {Springer International Publishing},\ \bibinfo {year} {2019})\ pp.\ \bibinfo {pages} {25--65}\BibitemShut {NoStop}%
\bibitem [{\citenamefont {Gholami~Mahmoodabadi}\ \emph {et~al.}(2020)\citenamefont {Gholami~Mahmoodabadi}, \citenamefont {Taylor}, \citenamefont {Kaller}, \citenamefont {Spindler}, \citenamefont {Mazaheri}, \citenamefont {Kasaian},\ and\ \citenamefont {Sandoghdar}}]{gholami2020point}%
  \BibitemOpen
  \bibfield  {author} {\bibinfo {author} {\bibfnamefont {R.}~\bibnamefont {Gholami~Mahmoodabadi}}, \bibinfo {author} {\bibfnamefont {R.~W.}\ \bibnamefont {Taylor}}, \bibinfo {author} {\bibfnamefont {M.}~\bibnamefont {Kaller}}, \bibinfo {author} {\bibfnamefont {S.}~\bibnamefont {Spindler}}, \bibinfo {author} {\bibfnamefont {M.}~\bibnamefont {Mazaheri}}, \bibinfo {author} {\bibfnamefont {K.}~\bibnamefont {Kasaian}},\ and\ \bibinfo {author} {\bibfnamefont {V.}~\bibnamefont {Sandoghdar}},\ }\href {https://doi.org/10.1364/OE.401374} {\bibfield  {journal} {\bibinfo  {journal} {Opt. Express}\ }\textbf {\bibinfo {volume} {28}},\ \bibinfo {pages} {25969} (\bibinfo {year} {2020})}\BibitemShut {NoStop}%
\bibitem [{\citenamefont {Dong}\ \emph {et~al.}(2021)\citenamefont {Dong}, \citenamefont {Maestre}, \citenamefont {Conrad-Billroth},\ and\ \citenamefont {Juffmann}}]{dong2021}%
  \BibitemOpen
  \bibfield  {author} {\bibinfo {author} {\bibfnamefont {J.}~\bibnamefont {Dong}}, \bibinfo {author} {\bibfnamefont {D.}~\bibnamefont {Maestre}}, \bibinfo {author} {\bibfnamefont {C.}~\bibnamefont {Conrad-Billroth}},\ and\ \bibinfo {author} {\bibfnamefont {T.}~\bibnamefont {Juffmann}},\ }\bibfield  {journal} {\bibinfo  {journal} {J. Phys. D: Appl. Phys.}\ }\textbf {\bibinfo {volume} {54}},\ \href {https://doi.org/10.1088/1361-6463/ac0f22} {10.1088/1361-6463/ac0f22} (\bibinfo {year} {2021})\BibitemShut {NoStop}%
\bibitem [{\citenamefont {Ruh}\ \emph {et~al.}(2018)\citenamefont {Ruh}, \citenamefont {Mutschler}, \citenamefont {Michelbach},\ and\ \citenamefont {Rohrbach}}]{ruh2018superior}%
  \BibitemOpen
  \bibfield  {author} {\bibinfo {author} {\bibfnamefont {D.}~\bibnamefont {Ruh}}, \bibinfo {author} {\bibfnamefont {J.}~\bibnamefont {Mutschler}}, \bibinfo {author} {\bibfnamefont {M.}~\bibnamefont {Michelbach}},\ and\ \bibinfo {author} {\bibfnamefont {A.}~\bibnamefont {Rohrbach}},\ }\href {https://doi.org/10.1364/OPTICA.5.001371} {\bibfield  {journal} {\bibinfo  {journal} {Optica}\ }\textbf {\bibinfo {volume} {5}},\ \bibinfo {pages} {1371} (\bibinfo {year} {2018})}\BibitemShut {NoStop}%
\bibitem [{\citenamefont {J{\"u}nger}\ \emph {et~al.}(2022)\citenamefont {J{\"u}nger}, \citenamefont {Ruh}, \citenamefont {Strobel}, \citenamefont {Michiels}, \citenamefont {Huber}, \citenamefont {Brandel}, \citenamefont {Madl}, \citenamefont {Gavrilov}, \citenamefont {Mihlan}, \citenamefont {Daller} \emph {et~al.}}]{junger2022100}%
  \BibitemOpen
  \bibfield  {author} {\bibinfo {author} {\bibfnamefont {F.}~\bibnamefont {J{\"u}nger}}, \bibinfo {author} {\bibfnamefont {D.}~\bibnamefont {Ruh}}, \bibinfo {author} {\bibfnamefont {D.}~\bibnamefont {Strobel}}, \bibinfo {author} {\bibfnamefont {R.}~\bibnamefont {Michiels}}, \bibinfo {author} {\bibfnamefont {D.}~\bibnamefont {Huber}}, \bibinfo {author} {\bibfnamefont {A.}~\bibnamefont {Brandel}}, \bibinfo {author} {\bibfnamefont {J.}~\bibnamefont {Madl}}, \bibinfo {author} {\bibfnamefont {A.}~\bibnamefont {Gavrilov}}, \bibinfo {author} {\bibfnamefont {M.}~\bibnamefont {Mihlan}}, \bibinfo {author} {\bibfnamefont {C.~C.}\ \bibnamefont {Daller}}, \emph {et~al.},\ }\href {https://doi.org/10.1038/s41467-022-29091-0} {\bibfield  {journal} {\bibinfo  {journal} {Nat. Commun.}\ }\textbf {\bibinfo {volume} {13}},\ \bibinfo {pages} {1758} (\bibinfo {year} {2022})}\BibitemShut {NoStop}%
\bibitem [{\citenamefont {Iqbal}\ \emph {et~al.}(2025)\citenamefont {Iqbal}, \citenamefont {Thiele}, \citenamefont {Pfitzner},\ and\ \citenamefont {Kukura}}]{iqbal2025enhanced}%
  \BibitemOpen
  \bibfield  {author} {\bibinfo {author} {\bibfnamefont {K.}~\bibnamefont {Iqbal}}, \bibinfo {author} {\bibfnamefont {J.~C.}\ \bibnamefont {Thiele}}, \bibinfo {author} {\bibfnamefont {E.}~\bibnamefont {Pfitzner}},\ and\ \bibinfo {author} {\bibfnamefont {P.}~\bibnamefont {Kukura}},\ }\href {https://doi.org/10.1021/acsphotonics.5c00123} {\bibfield  {journal} {\bibinfo  {journal} {ACS Photonics}\ }\textbf {\bibinfo {volume} {12}},\ \bibinfo {pages} {2647} (\bibinfo {year} {2025})}\BibitemShut {NoStop}%
\bibitem [{\citenamefont {Trees}(2001)}]{van2004detection}%
  \BibitemOpen
  \bibfield  {author} {\bibinfo {author} {\bibfnamefont {H.~V.}\ \bibnamefont {Trees}},\ }\bibinfo {title} {Classical detection and estimation theory},\ in\ \href {https://doi.org/https://doi.org/10.1002/0471221082.ch2} {\emph {\bibinfo {booktitle} {Detection, Estimation, and Modulation Theory}}}\ (\bibinfo  {publisher} {John Wiley \& Sons, Ltd},\ \bibinfo {year} {2001})\ Chap.~\bibinfo {chapter} {2}, pp.\ \bibinfo {pages} {19--165}\BibitemShut {NoStop}%
\bibitem [{\citenamefont {Backlund}\ \emph {et~al.}(2018)\citenamefont {Backlund}, \citenamefont {Shechtman},\ and\ \citenamefont {Walsworth}}]{backlund2018fundamental}%
  \BibitemOpen
  \bibfield  {author} {\bibinfo {author} {\bibfnamefont {M.~P.}\ \bibnamefont {Backlund}}, \bibinfo {author} {\bibfnamefont {Y.}~\bibnamefont {Shechtman}},\ and\ \bibinfo {author} {\bibfnamefont {R.~L.}\ \bibnamefont {Walsworth}},\ }\href {https://doi.org/10.1103/PhysRevLett.121.023904} {\bibfield  {journal} {\bibinfo  {journal} {Phys. Rev. Lett.}\ }\textbf {\bibinfo {volume} {121}},\ \bibinfo {pages} {023904} (\bibinfo {year} {2018})}\BibitemShut {NoStop}%
\bibitem [{\citenamefont {Bouchet}\ \emph {et~al.}(2021)\citenamefont {Bouchet}, \citenamefont {Rotter},\ and\ \citenamefont {Mosk}}]{bouchet2021maximum}%
  \BibitemOpen
  \bibfield  {author} {\bibinfo {author} {\bibfnamefont {D.}~\bibnamefont {Bouchet}}, \bibinfo {author} {\bibfnamefont {S.}~\bibnamefont {Rotter}},\ and\ \bibinfo {author} {\bibfnamefont {A.~P.}\ \bibnamefont {Mosk}},\ }\href {https://doi.org/10.1038/s41567-020-01137-4} {\bibfield  {journal} {\bibinfo  {journal} {Nat. Phys.}\ }\textbf {\bibinfo {volume} {17}},\ \bibinfo {pages} {564} (\bibinfo {year} {2021})}\BibitemShut {NoStop}%
\bibitem [{\citenamefont {Hitzelhammer}\ \emph {et~al.}(2024)\citenamefont {Hitzelhammer}, \citenamefont {Dostálová}, \citenamefont {Zykov}, \citenamefont {Platzer}, \citenamefont {Conrad-Billroth}, \citenamefont {Juffmann},\ and\ \citenamefont {Hohenester}}]{hitzelhammer24}%
  \BibitemOpen
  \bibfield  {author} {\bibinfo {author} {\bibfnamefont {F.}~\bibnamefont {Hitzelhammer}}, \bibinfo {author} {\bibfnamefont {A.}~\bibnamefont {Dostálová}}, \bibinfo {author} {\bibfnamefont {I.}~\bibnamefont {Zykov}}, \bibinfo {author} {\bibfnamefont {B.}~\bibnamefont {Platzer}}, \bibinfo {author} {\bibfnamefont {C.}~\bibnamefont {Conrad-Billroth}}, \bibinfo {author} {\bibfnamefont {T.}~\bibnamefont {Juffmann}},\ and\ \bibinfo {author} {\bibfnamefont {U.}~\bibnamefont {Hohenester}},\ }\href {https://doi.org/10.1021/acsphotonics.4c00621} {\bibfield  {journal} {\bibinfo  {journal} {ACS Photonics}\ }\textbf {\bibinfo {volume} {11}},\ \bibinfo {pages} {2745} (\bibinfo {year} {2024})}\BibitemShut {NoStop}%
\bibitem [{\citenamefont {Hüpfl}\ \emph {et~al.}(2024)\citenamefont {Hüpfl}, \citenamefont {Russo}, \citenamefont {Rachbauer}, \citenamefont {Bouchet}, \citenamefont {Lu}, \citenamefont {Kuhl},\ and\ \citenamefont {Rotter}}]{hupfl_continuity_2024}%
  \BibitemOpen
  \bibfield  {author} {\bibinfo {author} {\bibfnamefont {J.}~\bibnamefont {Hüpfl}}, \bibinfo {author} {\bibfnamefont {F.}~\bibnamefont {Russo}}, \bibinfo {author} {\bibfnamefont {L.~M.}\ \bibnamefont {Rachbauer}}, \bibinfo {author} {\bibfnamefont {D.}~\bibnamefont {Bouchet}}, \bibinfo {author} {\bibfnamefont {J.}~\bibnamefont {Lu}}, \bibinfo {author} {\bibfnamefont {U.}~\bibnamefont {Kuhl}},\ and\ \bibinfo {author} {\bibfnamefont {S.}~\bibnamefont {Rotter}},\ }\href {https://doi.org/10.1038/s41567-024-02519-8} {\bibfield  {journal} {\bibinfo  {journal} {Nat. Phys.}\ }\textbf {\bibinfo {volume} {20}},\ \bibinfo {pages} {1294} (\bibinfo {year} {2024})}\BibitemShut {NoStop}%
\bibitem [{\citenamefont {Ginsberg}\ \emph {et~al.}(2025)\citenamefont {Ginsberg}, \citenamefont {Hsieh}, \citenamefont {Kukura}, \citenamefont {Piliarik},\ and\ \citenamefont {Sandoghdar}}]{ginsberg2025interferometric}%
  \BibitemOpen
  \bibfield  {author} {\bibinfo {author} {\bibfnamefont {N.~S.}\ \bibnamefont {Ginsberg}}, \bibinfo {author} {\bibfnamefont {C.-L.}\ \bibnamefont {Hsieh}}, \bibinfo {author} {\bibfnamefont {P.}~\bibnamefont {Kukura}}, \bibinfo {author} {\bibfnamefont {M.}~\bibnamefont {Piliarik}},\ and\ \bibinfo {author} {\bibfnamefont {V.}~\bibnamefont {Sandoghdar}},\ }\href {https://doi.org/10.1038/s43586-025-00391-1} {\bibfield  {journal} {\bibinfo  {journal} {Nat Rev Methods Primers}\ }\textbf {\bibinfo {volume} {5}},\ \bibinfo {pages} {23} (\bibinfo {year} {2025})}\BibitemShut {NoStop}%
\bibitem [{\citenamefont {Hohenester}\ \emph {et~al.}(2022)\citenamefont {Hohenester}, \citenamefont {Reichelt},\ and\ \citenamefont {Unger}}]{hohenester2022nanophotonic}%
  \BibitemOpen
  \bibfield  {author} {\bibinfo {author} {\bibfnamefont {U.}~\bibnamefont {Hohenester}}, \bibinfo {author} {\bibfnamefont {N.}~\bibnamefont {Reichelt}},\ and\ \bibinfo {author} {\bibfnamefont {G.}~\bibnamefont {Unger}},\ }\href {https://doi.org/10.1016/j.cpc.2022.108337} {\bibfield  {journal} {\bibinfo  {journal} {Comp. Phys. Commun.}\ }\textbf {\bibinfo {volume} {276}},\ \bibinfo {pages} {108337} (\bibinfo {year} {2022})}\BibitemShut {NoStop}%
\bibitem [{\citenamefont {Hohenester}(2024)}]{hohenester2024nanophotonic}%
  \BibitemOpen
  \bibfield  {author} {\bibinfo {author} {\bibfnamefont {U.}~\bibnamefont {Hohenester}},\ }\href {https://doi.org/https://doi.org/10.1016/j.cpc.2023.108949} {\bibfield  {journal} {\bibinfo  {journal} {Comp. Phys. Commun.}\ }\textbf {\bibinfo {volume} {294}},\ \bibinfo {pages} {108949} (\bibinfo {year} {2024})}\BibitemShut {NoStop}%
\bibitem [{\citenamefont {Liu}\ \emph {et~al.}(2025)\citenamefont {Liu}, \citenamefont {Stergiopoulou}, \citenamefont {Chuah}, \citenamefont {Unser}, \citenamefont {Sage},\ and\ \citenamefont {Dong}}]{liu2025revisiting}%
  \BibitemOpen
  \bibfield  {author} {\bibinfo {author} {\bibfnamefont {Y.}~\bibnamefont {Liu}}, \bibinfo {author} {\bibfnamefont {V.}~\bibnamefont {Stergiopoulou}}, \bibinfo {author} {\bibfnamefont {J.}~\bibnamefont {Chuah}}, \bibinfo {author} {\bibfnamefont {M.}~\bibnamefont {Unser}}, \bibinfo {author} {\bibfnamefont {D.}~\bibnamefont {Sage}},\ and\ \bibinfo {author} {\bibfnamefont {J.}~\bibnamefont {Dong}},\ }\bibfield  {journal} {\bibinfo  {journal} {arXiv preprint arXiv:2502.03170}\ }\href {https://doi.org/10.48550/arXiv.2502.03170} {10.48550/arXiv.2502.03170} (\bibinfo {year} {2025})\BibitemShut {NoStop}%
\bibitem [{\citenamefont {Novotny}\ and\ \citenamefont {Hecht}(2012)}]{novotny2012principles}%
  \BibitemOpen
  \bibfield  {author} {\bibinfo {author} {\bibfnamefont {L.}~\bibnamefont {Novotny}}\ and\ \bibinfo {author} {\bibfnamefont {B.}~\bibnamefont {Hecht}},\ }\href {https://doi.org/10.1017/CBO9780511794193} {\emph {\bibinfo {title} {Principles of Nano-Optics}}},\ \bibinfo {edition} {2nd}\ ed.\ (\bibinfo  {publisher} {Cambridge University Press},\ \bibinfo {year} {2012})\BibitemShut {NoStop}%
\bibitem [{\citenamefont {Foley}\ \emph {et~al.}(2021)\citenamefont {Foley}, \citenamefont {Kushwah}, \citenamefont {Young},\ and\ \citenamefont {Kukura}}]{foley_mass_2021}%
  \BibitemOpen
  \bibfield  {author} {\bibinfo {author} {\bibfnamefont {E.~D.~B.}\ \bibnamefont {Foley}}, \bibinfo {author} {\bibfnamefont {M.~S.}\ \bibnamefont {Kushwah}}, \bibinfo {author} {\bibfnamefont {G.}~\bibnamefont {Young}},\ and\ \bibinfo {author} {\bibfnamefont {P.}~\bibnamefont {Kukura}},\ }\href {https://doi.org/10.1038/s41592-021-01261-w} {\bibfield  {journal} {\bibinfo  {journal} {Nat. Methods}\ }\textbf {\bibinfo {volume} {18}},\ \bibinfo {pages} {1247} (\bibinfo {year} {2021})}\BibitemShut {NoStop}%
\bibitem [{\citenamefont {Heermann}\ \emph {et~al.}(2021)\citenamefont {Heermann}, \citenamefont {Steiert}, \citenamefont {Ramm}, \citenamefont {Hundt},\ and\ \citenamefont {Schwille}}]{heermann_mass-sensitive_2021}%
  \BibitemOpen
  \bibfield  {author} {\bibinfo {author} {\bibfnamefont {T.}~\bibnamefont {Heermann}}, \bibinfo {author} {\bibfnamefont {F.}~\bibnamefont {Steiert}}, \bibinfo {author} {\bibfnamefont {B.}~\bibnamefont {Ramm}}, \bibinfo {author} {\bibfnamefont {N.}~\bibnamefont {Hundt}},\ and\ \bibinfo {author} {\bibfnamefont {P.}~\bibnamefont {Schwille}},\ }\href {https://doi.org/10.1038/s41592-021-01260-x} {\bibfield  {journal} {\bibinfo  {journal} {Nat. Methods}\ }\textbf {\bibinfo {volume} {18}},\ \bibinfo {pages} {1239} (\bibinfo {year} {2021})}\BibitemShut {NoStop}%
\bibitem [{\citenamefont {Bujak}\ \emph {et~al.}(2021)\citenamefont {Bujak}, \citenamefont {Holanová}, \citenamefont {García~Marín}, \citenamefont {Henrichs}, \citenamefont {Barvík}, \citenamefont {Braun}, \citenamefont {Lánský},\ and\ \citenamefont {Piliarik}}]{bujak2021}%
  \BibitemOpen
  \bibfield  {author} {\bibinfo {author} {\bibfnamefont {Å.}~\bibnamefont {Bujak}}, \bibinfo {author} {\bibfnamefont {K.}~\bibnamefont {Holanová}}, \bibinfo {author} {\bibfnamefont {A.}~\bibnamefont {García~Marín}}, \bibinfo {author} {\bibfnamefont {V.}~\bibnamefont {Henrichs}}, \bibinfo {author} {\bibfnamefont {I.}~\bibnamefont {Barvík}}, \bibinfo {author} {\bibfnamefont {M.}~\bibnamefont {Braun}}, \bibinfo {author} {\bibfnamefont {Z.}~\bibnamefont {Lánský}},\ and\ \bibinfo {author} {\bibfnamefont {M.}~\bibnamefont {Piliarik}},\ }\href {https://doi.org/10.1002/smtd.202100370} {\bibfield  {journal} {\bibinfo  {journal} {Small Methods}\ }\textbf {\bibinfo {volume} {5}},\ \bibinfo {pages} {2100370} (\bibinfo {year} {2021})}\BibitemShut {NoStop}%
\bibitem [{\citenamefont {Lipovsky}\ \emph {et~al.}(2010)\citenamefont {Lipovsky}, \citenamefont {Nitzan}, \citenamefont {Gedanken},\ and\ \citenamefont {Lubart}}]{lipovsky_visible_2010}%
  \BibitemOpen
  \bibfield  {author} {\bibinfo {author} {\bibfnamefont {A.}~\bibnamefont {Lipovsky}}, \bibinfo {author} {\bibfnamefont {Y.}~\bibnamefont {Nitzan}}, \bibinfo {author} {\bibfnamefont {A.}~\bibnamefont {Gedanken}},\ and\ \bibinfo {author} {\bibfnamefont {R.}~\bibnamefont {Lubart}},\ }\href {https://doi.org/10.1002/lsm.20948} {\bibfield  {journal} {\bibinfo  {journal} {Lasers Surg. Med.}\ }\textbf {\bibinfo {volume} {42}},\ \bibinfo {pages} {467} (\bibinfo {year} {2010})}\BibitemShut {NoStop}%
\bibitem [{\citenamefont {Abana}\ \emph {et~al.}(2017)\citenamefont {Abana}, \citenamefont {Brannon}, \citenamefont {Ebbott}, \citenamefont {Dunigan}, \citenamefont {Guckes}, \citenamefont {Fuseini}, \citenamefont {Powers}, \citenamefont {Rogers},\ and\ \citenamefont {Hadjifrangiskou}}]{abana_characterization_2017}%
  \BibitemOpen
  \bibfield  {author} {\bibinfo {author} {\bibfnamefont {C.~M.}\ \bibnamefont {Abana}}, \bibinfo {author} {\bibfnamefont {J.~R.}\ \bibnamefont {Brannon}}, \bibinfo {author} {\bibfnamefont {R.~A.}\ \bibnamefont {Ebbott}}, \bibinfo {author} {\bibfnamefont {T.~L.}\ \bibnamefont {Dunigan}}, \bibinfo {author} {\bibfnamefont {K.~R.}\ \bibnamefont {Guckes}}, \bibinfo {author} {\bibfnamefont {H.}~\bibnamefont {Fuseini}}, \bibinfo {author} {\bibfnamefont {J.}~\bibnamefont {Powers}}, \bibinfo {author} {\bibfnamefont {B.~R.}\ \bibnamefont {Rogers}},\ and\ \bibinfo {author} {\bibfnamefont {M.}~\bibnamefont {Hadjifrangiskou}},\ }\href {https://doi.org/10.1002/mbo3.466} {\bibfield  {journal} {\bibinfo  {journal} {Microbiologyopen}\ }\textbf {\bibinfo {volume} {6}},\ \bibinfo {pages} {e00466} (\bibinfo {year} {2017})}\BibitemShut {NoStop}%
\bibitem [{\citenamefont {Cuny}\ \emph {et~al.}(2022)\citenamefont {Cuny}, \citenamefont {Schlottmann}, \citenamefont {Ewald}, \citenamefont {Pelet},\ and\ \citenamefont {Schmoller}}]{Cuny2022LiveCM}%
  \BibitemOpen
  \bibfield  {author} {\bibinfo {author} {\bibfnamefont {A.~P.}\ \bibnamefont {Cuny}}, \bibinfo {author} {\bibfnamefont {F.~P.}\ \bibnamefont {Schlottmann}}, \bibinfo {author} {\bibfnamefont {J.~C.}\ \bibnamefont {Ewald}}, \bibinfo {author} {\bibfnamefont {S.}~\bibnamefont {Pelet}},\ and\ \bibinfo {author} {\bibfnamefont {K.~M.}\ \bibnamefont {Schmoller}},\ }\bibfield  {journal} {\bibinfo  {journal} {Biophys. Rev.}\ }\textbf {\bibinfo {volume} {3}},\ \href {https://doi.org/10.1063/5.0082799} {10.1063/5.0082799} (\bibinfo {year} {2022})\BibitemShut {NoStop}%
\bibitem [{\citenamefont {Schmid}\ \emph {et~al.}(2014)\citenamefont {Schmid}, \citenamefont {Andrae},\ and\ \citenamefont {Manley}}]{schmid2014plasmonic}%
  \BibitemOpen
  \bibfield  {author} {\bibinfo {author} {\bibfnamefont {M.}~\bibnamefont {Schmid}}, \bibinfo {author} {\bibfnamefont {P.}~\bibnamefont {Andrae}},\ and\ \bibinfo {author} {\bibfnamefont {P.}~\bibnamefont {Manley}},\ }\href {https://doi.org/10.1186/1556-276X-9-50} {\bibfield  {journal} {\bibinfo  {journal} {Nanoscale Res. Lett.}\ }\textbf {\bibinfo {volume} {9}},\ \bibinfo {pages} {50} (\bibinfo {year} {2014})}\BibitemShut {NoStop}%
\bibitem [{\citenamefont {Gonzalez-Ballestero}\ \emph {et~al.}(2021)\citenamefont {Gonzalez-Ballestero}, \citenamefont {Aspelmeyer}, \citenamefont {Novotny}, \citenamefont {Quidant},\ and\ \citenamefont {Romero-Isart}}]{levitation21}%
  \BibitemOpen
  \bibfield  {author} {\bibinfo {author} {\bibfnamefont {C.}~\bibnamefont {Gonzalez-Ballestero}}, \bibinfo {author} {\bibfnamefont {M.}~\bibnamefont {Aspelmeyer}}, \bibinfo {author} {\bibfnamefont {L.}~\bibnamefont {Novotny}}, \bibinfo {author} {\bibfnamefont {R.}~\bibnamefont {Quidant}},\ and\ \bibinfo {author} {\bibfnamefont {O.}~\bibnamefont {Romero-Isart}},\ }\href {https://doi.org/10.1126/science.abg3027} {\bibfield  {journal} {\bibinfo  {journal} {Science}\ }\textbf {\bibinfo {volume} {374}},\ \bibinfo {pages} {eabg3027} (\bibinfo {year} {2021})}\BibitemShut {NoStop}%
\bibitem [{\citenamefont {Ketterle}\ \emph {et~al.}(1999)\citenamefont {Ketterle}, \citenamefont {Durfee},\ and\ \citenamefont {Stamper-Kurn}}]{ketterle1999making}%
  \BibitemOpen
  \bibfield  {author} {\bibinfo {author} {\bibfnamefont {W.}~\bibnamefont {Ketterle}}, \bibinfo {author} {\bibfnamefont {D.~S.}\ \bibnamefont {Durfee}},\ and\ \bibinfo {author} {\bibfnamefont {D.~M.}\ \bibnamefont {Stamper-Kurn}},\ }\href {https://doi.org/10.3254/978-1-61499-225-7-67} {\bibfield  {journal} {\bibinfo  {journal} {IOS press}\ }\textbf {\bibinfo {volume} {140}},\ \bibinfo {pages} {67} (\bibinfo {year} {1999})}\BibitemShut {NoStop}%
\bibitem [{\citenamefont {Shechtman}\ \emph {et~al.}(2014)\citenamefont {Shechtman}, \citenamefont {Sahl}, \citenamefont {Backer},\ and\ \citenamefont {Moerner}}]{shechtman2014}%
  \BibitemOpen
  \bibfield  {author} {\bibinfo {author} {\bibfnamefont {Y.}~\bibnamefont {Shechtman}}, \bibinfo {author} {\bibfnamefont {S.~J.}\ \bibnamefont {Sahl}}, \bibinfo {author} {\bibfnamefont {A.~S.}\ \bibnamefont {Backer}},\ and\ \bibinfo {author} {\bibfnamefont {W.~E.}\ \bibnamefont {Moerner}},\ }\href {https://doi.org/10.1103/PhysRevLett.113.133902} {\bibfield  {journal} {\bibinfo  {journal} {Phys. Rev. Lett.}\ }\textbf {\bibinfo {volume} {113}},\ \bibinfo {pages} {133902} (\bibinfo {year} {2014})}\BibitemShut {NoStop}%
\bibitem [{\citenamefont {Chao}\ \emph {et~al.}(2016)\citenamefont {Chao}, \citenamefont {Ward},\ and\ \citenamefont {Ober}}]{Chao2016}%
  \BibitemOpen
  \bibfield  {author} {\bibinfo {author} {\bibfnamefont {J.}~\bibnamefont {Chao}}, \bibinfo {author} {\bibfnamefont {E.~S.}\ \bibnamefont {Ward}},\ and\ \bibinfo {author} {\bibfnamefont {R.~J.}\ \bibnamefont {Ober}},\ }\href {https://doi.org/10.1364/JOSAA.33.000B36} {\bibfield  {journal} {\bibinfo  {journal} {J. Opt. Soc. Am. A}\ }\textbf {\bibinfo {volume} {33}},\ \bibinfo {pages} {B36} (\bibinfo {year} {2016})}\BibitemShut {NoStop}%
\bibitem [{\citenamefont {Juffmann}\ \emph {et~al.}(2025)\citenamefont {Juffmann}, \citenamefont {Nimmrichter}, \citenamefont {Balzarotti},\ and\ \citenamefont {Dong}}]{juffmann2025quantum}%
  \BibitemOpen
  \bibfield  {author} {\bibinfo {author} {\bibfnamefont {T.}~\bibnamefont {Juffmann}}, \bibinfo {author} {\bibfnamefont {S.}~\bibnamefont {Nimmrichter}}, \bibinfo {author} {\bibfnamefont {F.}~\bibnamefont {Balzarotti}},\ and\ \bibinfo {author} {\bibfnamefont {J.}~\bibnamefont {Dong}},\ }\href {https://doi.org/10.1051/photon/202513158} {\bibfield  {journal} {\bibinfo  {journal} {Photoniques}\ }\textbf {\bibinfo {volume} {131}},\ \bibinfo {pages} {58} (\bibinfo {year} {2025})}\BibitemShut {NoStop}%
\end{thebibliography}%

\end{document}